\documentclass[sigconf,screen]{acmart}
\usepackage{doi}
\usepackage[title]{appendix}
\usepackage[utf8]{inputenc}
\usepackage[T1]{fontenc}
\usepackage{microtype}
\usepackage{soul, color, xcolor}
\usepackage{amsmath,amsfonts}
\usepackage{algorithmicx}
\usepackage{graphicx}
\usepackage{textcomp}
\usepackage{xcolor}
\usepackage{siunitx}
\usepackage{comment}
\usepackage{tcolorbox}
\usepackage{listings}
\usepackage{xspace}
\usepackage{float}
\usepackage{graphicx}
\usepackage{subfigure}
\usepackage{pifont}
\usepackage{threeparttable,booktabs,multirow,lscape}
\usepackage{multirow}
\usepackage{adjustbox}
\usepackage{enumerate}
\usepackage{listings}
\usepackage{tabularx}
\usepackage[ruled,vlined]{algorithm2e}
\usepackage{amsmath}
\usepackage{environ}
\usepackage{tikz}
\usepackage{color}
\usepackage{xcolor}

 \usepackage{amssymb}
\usepackage{enumitem}
\usepackage{rotating}
\usepackage{threeparttable}
\usepackage{url}
\usepackage{hyperref}
\usepackage{caption}
\usepackage{balance}
\AtBeginDocument{%
  \providecommand\BibTeX{{%
    \normalfont B\kern-0.5em{\scshape i\kern-0.25em b}\kern-0.8em\TeX}}}

\newcommand{\tab}{\hspace*{1em}}

\newcommand{\ie}{\emph{i.e.}\xspace}
\newcommand{\eg}{\emph{e.g.}\xspace}

\newcommand{\Eg}{\emph{E.g.}\xspace}
\newcommand{\system}{DeUEDroid}
\newcommand{\target}{UEware}
\newcommand{\legit}{normal app}
\newcommand{\leg}{normal}

\newcommand{\sg}{UTG}
\soulregister{\cite}7 % 
\soulregister{\citep}7 % 
\soulregister{\citet}7 % 
\soulregister{\ref}7 % 
\soulregister{\pageref}7 % 

\def\BibTeX{{\rm B\kern-.05em{\sc i\kern-.025em b}\kern-.08em
    T\kern-.1667em\lower.7ex\hbox{E}\kern-.125emX}}

\setcopyright{acmlicensed}
\acmPrice{15.00}
\acmDOI{10.1145/3597926.3598051}
\acmYear{2023}
\copyrightyear{2023}
\acmSubmissionID{issta23main-p73-p}
\acmISBN{979-8-4007-0221-1/23/07}
\acmConference[ISSTA '23]{Proceedings of the 32nd ACM SIGSOFT International Symposium on Software Testing and Analysis}{July 17--21, 2023}{Seattle, WA, USA}
\acmBooktitle{Proceedings of the 32nd ACM SIGSOFT International Symposium on Software Testing and Analysis (ISSTA '23), July 17--21, 2023, Seattle, WA, USA}
\received{2023-02-16}
\received[accepted]{2023-05-03}

\begin{document}

\title{DeUEDroid: Detecting Underground Economy Apps Based on UTG Similarity}


\settopmatter{authorsperrow=4} %make the template consider one author per row

% \author{
% {\rm Zhuo~Chen$^{1,2}$, Jie~Liu$^{2}$, Yubo~Hu$^{3}$, Lei~Wu$^{1,4}$*, Yajin~Zhou$^{1,4}$, Yiling~He$^{1}$, Xianhao~Liao$^{2}$}\\
% {\rm Ke~Wang$^{2}$, Jinku~Li$^{3}$, Zhan~Qin$^{1,4}$}\\
% $^{1}$Zhejiang University
% $^{2}$Ant Group 
% $^{3}$Xidian University \\
% $^{4}$Key Laboratory of Blockchain and Cyberspace Governance of Zhejiang Province \\
% }

\author{Zhuo~Chen}
\email{hypothesiser.hypo@zju.edu.cn}
\affiliation{
\institution{Zhejiang University \\ \& Ant Group}
\country{China}
}

\author{Jie~Liu}
\email{qingyang.lj@antgroup.com}
\affiliation{
\institution{Ant Group}
\country{China}
}

\author{Yubo~Hu}
\email{yubohu@stu.xidian.edu.cn}
\affiliation{
\institution{Xidian University}
\country{China}
}

\author{Lei~Wu}
\email{lei_wu@zju.edu.cn}
\affiliation{
\institution{Zhejiang University}
\country{China}
}
\authornote{Corresponding author.}

\author{Yajin~Zhou}
\email{yajin_zhou@zju.edu.cn}
\affiliation{
\institution{Zhejiang University}
\country{China}
}

\author{Yiling~He}
\email{yilinghe@zju.edu.cn}
\affiliation{
\institution{Zhejiang University}
\country{China}
}

\author{Xianhao~Liao}
\email{xianhao.lxh@antgroup.com}
\affiliation{
\institution{Ant Group}
\country{China}
}

\author{Ke~Wang}
\email{kaywang.wk@antgroup.com}
\affiliation{
\institution{Ant Group}
\country{China}
}

\author{Jinku~Li}
\email{jkli@xidian.edu.cn}
\affiliation{
\institution{Xidian University}
\country{China}
}

\author{Zhan~Qin}
\email{qinzhan@zju.edu.can}
\affiliation{
\institution{Zhejiang University}
\country{China}
}

% \newcommand{\tsc}[1]{\textsuperscript{#1}} %shorthand for superscripts
% \author{Zhuo~Chen\tsc{1,2}, Jieliu\tsc{2}, Yubo~Hu\tsc{3}, Lei~Wu\tsc{1}}
% \author{Yajin~Zhou\tsc{1}, Yiling~He\tsc{1}, Xianhao~Liao\tsc{2}, Ke~Wang\tsc{2}} %to break line, start another author block
% \author{Jinku~Li\tsc{3}, Zhan~Qin\tsc{1}}
% \affiliation{
%   \institution{\vskip .2cm}%add some spacing if needed
%   \institution{1. Zhejiang University 2.Ant Group 3.Xidian University}
%   % \city{Hangzhou}
%   \country{China}
% }


\begin{abstract}

In recent years, the underground economy is proliferating in the mobile system. 
These underground economy apps (\textbf{UEware} for short) make profits from \textit{providing non-compliant services}, especially in sensitive areas (\eg, gambling, porn, loan). Unlike traditional malware, most of them (over 80\%) do not have malicious payloads.
Due to their unique characteristics, existing detection approaches cannot effectively and efficiently mitigate this emerging threat.
  
  To address this problem, we propose a novel approach to effectively and efficiently detect \target{} by considering their UI transition graphs (UTGs).
  Based on the proposed approach, we design and implement a system, named \system{}, to perform the detection. 
  To evaluate \system{}, we collect $25,717$ apps and build up the first large-scale ground-truth dataset ($1,700$ apps) of \target{}. 
  The evaluation result based on the ground-truth dataset shows that \system{} can cover new UI features and statically construct precise \sg{}. It achieves \num{98.22}\% detection F1-score and \num{98.97}\% classification accuracy, a significantly better performance than the traditional approaches. 
  The evaluation result involving $24,017$ apps demonstrates the effectiveness and efficiency of \target{} detection in real-world scenarios.
  Furthermore, the result also reveals that \target{} are prevalent, \ie, 54\% apps in the wild and 11\% apps in the app stores are \target{}.
  Our work sheds light on the future work of analyzing and detecting \target{}.
  To engage the community, we have made our prototype system and the dataset available online.
\end{abstract}



% \ccsdesc[500]{Security and privacy~Software security engineering}
% \ccsdesc{Economics of security and privacy}


\renewcommand{\shortauthors}{Z. Chen, J. Liu, Y. Hu, L. Wu, Y. Zhou, Y. He, X. Liao, K. Wang, J. Li, Z. Qin}
% \renewcommand{\authors}{\shortauthors}

\maketitle

\renewcommand{\thefootnote}{\fnsymbol{footnote}}

\renewcommand{\thefootnote}{\arabic{footnote}}

\section{Introduction}
\label{sec:intro}
%Mobile apps have been an indispensable part of our daily life, from online shopping, entertainment, and even financial business~\cite{economyapp}.
Mobile apps have been an indispensable part of our daily life, even in some sensitive areas, such as adult content, financial business~\cite{economyapp}, and online gambling.
However, due to the huge economic benefits, apps that serve the underground economy are prevalent in these areas nowadays~\cite{chinaGambling,hu2018dating,loan}, and thereby lead to serious damages. 
For example, the underground porn apps caused more than \$\num{304} million losses in 2020~\cite{BBC}.
Meanwhile, underground gambling apps made more than \$1 billion revenue in Malaysia in 2021~\cite{AsiaScam}. 

\begin{figure}[!t]
\centering
	\includegraphics[width=.45\textwidth]{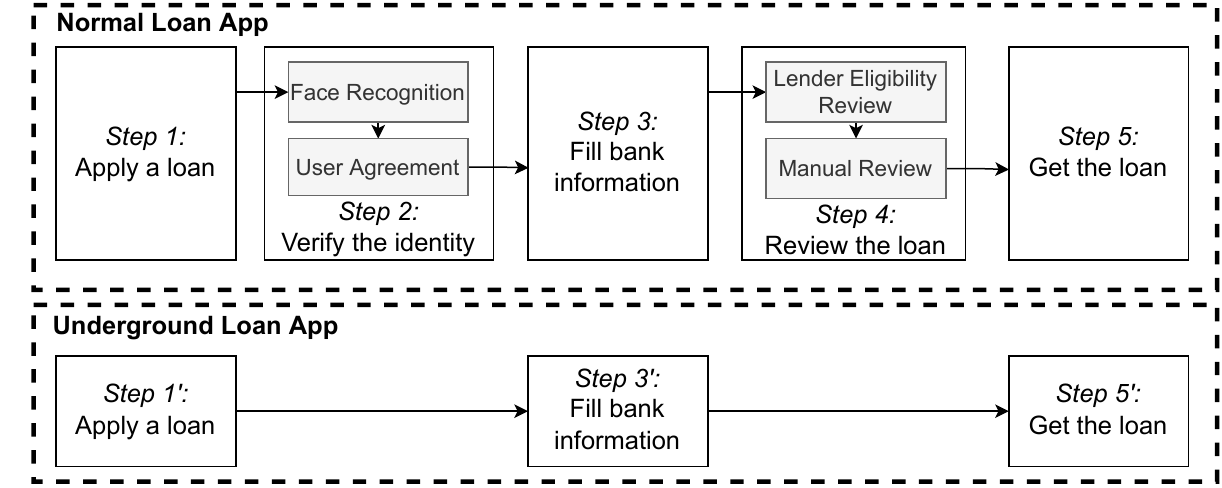}
	%\caption{The definition of \target{} compared to malware.}
         \caption{An example of a normal loan app vs. an underground loan app. The underground loan app only provides a single usury service (Steps 1',3',5') without verifying the identity (Step 2) and reviewing the loan (Step 4).}
	\label{fig:def}
 % \vspace{-1em}
\end{figure}

Unlike traditional malware, these underground apps make profits by \textit{providing non-compliant services}, especially in sensitive areas (\eg, gambling, porn, loan).
For better understanding, Figure~\ref{fig:def} shows a typical use scenario of a normal~\footnote{In this paper, apps that do not belong to \target{} are named as \textit{\legit{}s}.} loan app and an underground loan app. 
Unlike the normal app, the underground app does not provide UIs related to \textit{terms and conditions}, including identity verification and user (or privacy policy) agreement. Such a phenomenon is probably because these apps want to avoid  supervision and let the victims make a quick decision without thinking~\cite{GooglePay}.
What's more, our investigation shows that most of these underground apps (over 80\%) do not have malicious payloads (or serve advertisements).
 
In this work, we call these apps \textit{underground economy apps} (or \textbf{UEware} for short), and define them as \textit{apps that serve the underground economy by providing non-compliant services}~\footnote{For a given sensitive service, the regulations may vary in different countries/regions, \eg, online gambling is prohibited in Saudi Arabia, while it is legal in some states of the United States~\cite{gamblingUS}. The impact is discussed in Section~\ref{sec:discussion}.}. %In this paper, we call such apps \textbf{\target{}} for short. 
Obviously, the concept is orthogonal to that of malware.

The proliferation of UEware has become a widespread issue, particularly on platforms that lack adequate supervision, such as social media and third-party app markets. 
To mitigate this threat, authoritative agencies and app download platforms proactively seek effective solutions.
However, to our best knowledge, only a few studies~\cite{muscanell2014weapons,gao2021demystifying, munyendo2022desperate,chen2021lifting,hongGamblnig2022} focus on a special sort of \target{} (\ie, gambling scam, loan scam apps), and provide some recommendations to users and authorities. 
Nonetheless, none of these works propose feasible approaches which are capable of effectively detecting \target{}. Apart from the effectiveness, efficiency is also significant due to the huge amount of apps~\footnote{\Eg, there were over two million new apps released on GooglePlay in 2021~\cite{appadd}.} that need to be handled.

Unfortunately, existing dynamic and static approaches that are used to detect malware cannot be applied to detect \target{}.
On one hand, the dynamic approaches are inefficient to perform the large-scale detection due to the inherent scalability limitation, i.e., they have to launch the apps to examine their behaviors~\cite{wong2016intellidroid} and fetch the screenshots~\cite{dong2018frauddroid}, thus consuming huge resources~\cite{li2017simidroid}. 
On the other hand, the static approaches are ineffective to detect \target{} due to its unique characteristics. Specifically, the features used by those detection approaches can be dived into malicious payload~\cite{enck2009lightweight,Wu2012DroidMatAM,yang2013appintent,liu2014decaf}, GUI content~\cite{yuan2019stealthy,sun2015droideagle}, and the Manifest information~\cite{faruki2015androsimilar,sebastian2020towards,li2018significant}.
However, these features are not feasible to \target{}: 1) \textit{most \target{} (over 80\%) do not have malicious payloads}, which inevitably makes the payload-based approaches ineffective; and 2) \textit{most \target{} take countermeasures}, which also makes the content-based approaches ineffective; and
3) the Manifest information cannot be used as a unique detection feature, since it is the basic information and can be arbitrarily customized in Android.

\noindent \textbf{Our approach.}\tab
In this paper, we propose a novel approach to efficiently and effectively detect Android \target{} by identifying the non-compliant services they provided.
Specifically, an app service is presented to the users through multiple user interfaces (UIs) in a specific order, which constitutes the \textit{UI transition graph} (\textbf{\sg{}}).

Our approach is based on the following two key observations.
First, \textit{the \sg{}s of \target{} are different from those of the \legit{}s}. Even \target{} imitate the UI of \legit{}s, their \sg{}s vary greatly.
As discussed earlier, \target{} do not provide UIs related to \textit{terms and conditions}, which will always be provided by \legit{}s.
Furthermore, \target{} are generally implemented to serve only one purpose (\eg, usury application), 
while \legit{}s are usually served for multiple services (\eg, financial derivatives purchasing and asset evaluation).
Second, \textit{the \sg{}s of the same type of \target{} are extremely similar, while the \sg{}s of different types of \target{} vary from each other}.
This is because different services share different \sg{}s. 
Based on these observations, we can use \sg{} as the key feature to perform the similarity-based detection.

\noindent \textbf{Challenges.} \tab
However, identifying \target{} based on \sg{} is a challenging task.  
First, \textit{it is difficult to directly leverage \sg{} to perform the \target{} detection}.
Although \sg{} has widely been adopted to assist the testing and security vetting in the previous studies~\cite{behrang2019test,tang2020all,wu2018sentinel}, to our best knowledge, it has never been used as a detection feature. This is due to the fact that \sg{} is expressed via a mixture of graph topology and unstructured GUI widget attributes.
As a result, their correlations cannot easily be applied as an effective representation. 
Second, \textit{it is not easy to statically build a precise \sg{}}. 
Some new UI features (\eg, WebView, fragment and navigation) are supported by Android and have been widely used in recent years.
However, (even the latest) tools/systems proposed by the previous studies~\cite{rountev2014static,kuznetsov2021all,octeau2015composite,liu2022promal,rountev2014static} cannot cover these new features. 
This inevitably leads to the imprecise \sg{} construction.

\noindent \textbf{This work.}\tab
In this study, we propose the first \sg{} based detection approach by addressing the above two challenges. 
For the first challenge, we propose a graph-embedding~\cite{hinton1993autoencoders,velickovic2019deep} based method to represent \sg{} by properly correlating multi-dimensional attributes. 
Specifically, the graph topology and the encoded attributes~\footnote{Different attributes will be encoded with different methods, see Section~\ref{sec:graphai}.} are aggregated and handled by applying the graph-embedding technique~\cite{hinton1993autoencoders,velickovic2019deep} to generate the \sg{} representation.
For the second challenge, we propose a taint-based construction method which can cover new features (\ie, identifying GUI elements and determining UI transitions) to build up a more precise \sg{}.

We have implemented a prototype, \system{}, to detect \target{}.
\system{} consists of three modules: \textit{(i) \sg{} builder}, which is used to construct the \sg{}.
This module consists of two sub-modules, \ie, \textit{GUI widget identification} and \textit{UI transition determination}.
\textit{(ii) \sg{} feature extractor}, which is responsible for extracting the \sg{} feature. 
\textit{(iii) \target{} detector.} 
The detector leverages the self-supervised learning to detect and classify \target{} based on \sg{} similarity.

We build up three datasets to evaluate \system{}. 
First, we use \num{15} source-available apps to evaluate the \sg{} construction accuracy.
The evaluation result suggests that \system{} is effective in  \sg{} construction, which can accurately identify imprint tokens of WebView (over 85\% F1-score) and transitions (92.3\% F1-score in total). Specifically, the transitions identified by \system{} achieves a significantly better F1-score than the State-Of-The-Art (SOTA) results (\ie, 35.5\% higher than IC3~\cite{octeau2015composite}, 67.7\% higher than GATOR~\cite{rountev2014static}).
Then, we build up the ground-truth \target{} dataset, which is the first large-scale ground-truth dataset ($1,700$ apps) of \target{}, to evaluate the effectiveness of \target{} detection and classification. 
The evaluation result shows that \system{} is effective in \target{} detection and classification (with \num{98.22}\% detection F1-score and \num{98.97}\% classification accuracy). Our system has a significantly better performance than the traditional approaches.
Finally, we perform a large-scale experiment on $24,017$ apps collected from both app stores and the wild, to measure the effectiveness and efficiency of \target{} detection in the real world.
The evaluation result shows that \target{} are prevalent, \ie, \num{54}\% apps in the wild and \num{11}\% apps in the app stores are \target{}.

\noindent \textbf{Contributions.}\tab
We make the following contributions:
\begin{itemize}[leftmargin=*]
    \item We proposed a novel approach to effectively and efficiently detect \target{} based on \sg{} similarity, including a novel technique that can cover new UI features and statically build precise \sg{}, and a novel algorithm that can efficiently embed \sg{} with different dimension attributes.
    \item We implemented a prototype named \system{}. The evaluation results demonstrate the effectiveness and efficiency of the system, with 98.22\% \target{} detection F1-score and 98.97\% classification accuracy. And it is capable of performing large-scale detection to mitigate the real-world threats.
    \item We found that \target{} are prevalent, \ie, \num{54}\% apps in the wild and \num{11}\% apps in the app stores are \target{}.
    \item We built up the first large-scale ground-truth \target{} dataset with $1,700$ apps. We have released our system and the dataset on the github~\cite{DeUEDroidtool} to engage the community.
\end{itemize}

\section{Background}
\label{sec:preliminary}

\begin{table*}[!tb]
\caption{The feature selection in previous static detection studies.}
% \vspace{-1.5em}
\centering
%\small

\resizebox{1\textwidth}{!}{
% \usepackage{multirow}
% \usepackage{multirow}
% \usepackage{multirow}
% \usepackage{multirow}
% \usepackage{multirow}
% \usepackage{multirow}

\begin{tabular}{|l|l|lllllllll|}
\hline
\multicolumn{1}{|c|}{\multirow{3}{*}{Static Detection Approach}} & \multirow{3}{*}{ Studies} & \multicolumn{9}{c|}{Feature Selection}  \\ \cline{3-11} 
\multicolumn{1}{|c|}{}  & & \multicolumn{4}{c|}{Malicious payload}  & \multicolumn{3}{c|}{GUI} & \multicolumn{2}{c|}{Manifest}  \\ \cline{3-11} 
\multicolumn{1}{|c|}{}  & & \multicolumn{1}{l|}{Bytecode}   & \multicolumn{1}{l|}{API}   & \multicolumn{1}{l|}{Method}  & \multicolumn{1}{l|}{Activity}   & \multicolumn{1}{l|}{Native widget}   & \multicolumn{1}{l|}{Web widget}  & \multicolumn{1}{l|}{GUI content}   & \multicolumn{1}{l|}{Dveloper signature}  & Permission   \\ \hline
\multirow{4}{*}{Signature Based}  & ViewDroid~\cite{zhang2014viewdroid}, MassVert~\cite{chen2015finding} & \multicolumn{1}{l|}{-}  & \multicolumn{1}{l|}{-}  & \multicolumn{1}{l|}{\ding{52}} & \multicolumn{1}{l|}{-}  & \multicolumn{1}{l|}{\ding{52}} & \multicolumn{1}{l|}{-}  & \multicolumn{1}{l|}{-}  & \multicolumn{1}{l|}{-}  & -  \\ \cline{2-11} 
 & DriodMoss~\cite{liu2006gplag} & \multicolumn{1}{l|}{\ding{52}} & \multicolumn{1}{l|}{-}  & \multicolumn{1}{l|}{-}  & \multicolumn{1}{l|}{-}  & \multicolumn{1}{l|}{-}  & \multicolumn{1}{l|}{-}  & \multicolumn{1}{l|}{-}  & \multicolumn{1}{l|}{-}  & -  \\ \cline{2-11} 
 & Retriever~\cite{sebastian2020towards}, AndroSimilar~\cite{faruki2015androsimilar} & \multicolumn{1}{l|}{-}  & \multicolumn{1}{l|}{-}  & \multicolumn{1}{l|}{-}  & \multicolumn{1}{l|}{-}  & \multicolumn{1}{l|}{-}  & \multicolumn{1}{l|}{-}  & \multicolumn{1}{l|}{-}  & \multicolumn{1}{l|}{\ding{52}} & -  \\ \cline{2-11} 
 & Kirin~\cite{enck2009lightweight}, DroidMat~\cite{Wu2012DroidMatAM}  & \multicolumn{1}{l|}{-}  & \multicolumn{1}{l|}{\ding{52}} & \multicolumn{1}{l|}{-}  & \multicolumn{1}{l|}{-}  & \multicolumn{1}{l|}{-}  & \multicolumn{1}{l|}{-}  & \multicolumn{1}{l|}{-}  & \multicolumn{1}{l|}{-}  & \ding{52} \\ \hline
\multirow{5}{*}{Machine Learning Based}  & Kim et al.~\cite{kim2018multimodal}  & \multicolumn{1}{l|}{-}  & \multicolumn{1}{l|}{\ding{52}} & \multicolumn{1}{l|}{\ding{52}} & \multicolumn{1}{l|}{-}  & \multicolumn{1}{l|}{-}  & \multicolumn{1}{l|}{-}  & \multicolumn{1}{l|}{-}  & \multicolumn{1}{l|}{-}  & \ding{52} \\ \cline{2-11} 
 %& DroidSIFT~\cite{yuan2014droid} & \multicolumn{1}{l|}{-}  & \multicolumn{1}{l|}{\ding{52}} & \multicolumn{1}{l|}{-}  & \multicolumn{1}{l|}{-}  & \multicolumn{1}{l|}{-}  & \multicolumn{1}{l|}{-}  & \multicolumn{1}{l|}{-}  & \multicolumn{1}{l|}{-}  & -  \\ \cline{2-11} 
 & DroidSIFT~\cite{yuan2014droid}, SIGPID~\cite{li2018significant}, Droid-Sec~\cite{ahmed2009using} & \multicolumn{1}{l|}{-}  & \multicolumn{1}{l|}{\ding{52}} & \multicolumn{1}{l|}{-}  & \multicolumn{1}{l|}{-}  & \multicolumn{1}{l|}{-}  & \multicolumn{1}{l|}{-}  & \multicolumn{1}{l|}{-}  & \multicolumn{1}{l|}{-}  & \ding{52} \\ \cline{2-11} 
& Tiger~\cite{chen2017mass}& \multicolumn{1}{l|}{-}  & \multicolumn{1}{l|}{-} & \multicolumn{1}{l|}{-}  & \multicolumn{1}{l|}{-}  & \multicolumn{1}{l|}{-}  & \multicolumn{1}{l|}{\ding{52}}  & \multicolumn{1}{l|}{-}  & \multicolumn{1}{l|}{-}  & - \\ \cline{2-11} 
  %\hline
& Malena~\cite{yuan2019stealthy}, Sun et al.~\cite{sun2015droideagle} & \multicolumn{1}{l|}{-}  & \multicolumn{1}{l|}{-} & \multicolumn{1}{l|}{-}  & \multicolumn{1}{l|}{-}  & \multicolumn{1}{l|}{-}  & \multicolumn{1}{l|}{-}  & \multicolumn{1}{l|}{\ding{52}}  & \multicolumn{1}{l|}{-}  & - \\ \cline{2-11} 

& Mamadroid~\cite{mariconti2016mamadroid} & \multicolumn{1}{l|}{-}  & \multicolumn{1}{l|}{\ding{52}} & \multicolumn{1}{l|}{-}  & \multicolumn{1}{l|}{-}  & \multicolumn{1}{l|}{-}  & \multicolumn{1}{l|}{-}  & \multicolumn{1}{l|}{-}  & \multicolumn{1}{l|}{-}  & - \\ \cline{2-11}

& Drebin~\cite{arp2014drebin} & \multicolumn{1}{l|}{-}  & \multicolumn{1}{l|}{\ding{52}} & \multicolumn{1}{l|}{-}  & \multicolumn{1}{l|}{\ding{52}}  & \multicolumn{1}{l|}{-}  & \multicolumn{1}{l|}{-}  & \multicolumn{1}{l|}{\ding{52}}  & \multicolumn{1}{l|}{-}  & \ding{52} 
%\\ \cline{2-11} 
  
% & \textbf{\system{}(Ours)} & \multicolumn{1}{l|}{-}  & \multicolumn{1}{l|}{-}  & \multicolumn{1}{l|}{-}  & \multicolumn{1}{l|}{\ding{52}} & \multicolumn{1}{l|}{\ding{52}} & \multicolumn{1}{l|}{\ding{52}} & \multicolumn{1}{l|}{-} & \multicolumn{1}{l|}{-}  & -  
\\ \hline
\end{tabular}

}

\label{tab:prewor}
\end{table*}

\subsection{Android User Interface}
\label{subsec:ui}
In an Android app, the activity is the fundamental component for drawing the GUI with which users can interact with~\cite{Google_Activity}. And since Android 3.0, fragments can also render the GUI.
A GUI consists of GUI widgets (\eg, button and textView) and layout models (\eg, Linearlayout) that describe how to arrange GUI widgets. These widgets and layout models inherit from the \textit{view} class, and are statically recorded in the XML file or added in code.
What's more, web widgets can launch a local browser to load a web page.

An android app consists of multiple activity/fragment windows for handling complex business requirements.
When some special conditions are triggered (\eg, click event \textit{onclick}), the app will transit from one window to another.
Specifically, these GUI transitions are triggered in multiple ways: \textit{(i)} triggered by explicitly invoking activity transition API, \eg, the \textit{StartActivity, StartActivityForResult}; \textit{(ii)} triggered by implicitly invoking conversions, \eg, Android Inter-Component Communication (\textit{ICC}); \textit{(iii)} triggered by invoking hardware event, \eg, \textit{BACK}; 
\textit{(iv)} triggered by fragment navigation, \eg, the \textit{NavController} manages fragment navigation within a \textit{NavHost}.
Note that fragment navigation extremely improves the flexibility of transitions, and has been widely used in recent years~\cite{Navigation}.

\subsection{Modeling of Android User Interface}
\label{subsec:sgdefination}
An android app is presented to the users  through multiple runtime-rendered user interfaces (UIs) in a specific order.
However, runtime rendered visuals cannot be restored in static analysis. To this end, the UI modeling studies~\cite{liu2022promal,chen2019storydroid,dong2018frauddroid} focus on expressing UI through static analysis in a limited time.
However, due to the rapid revolution of Android, there are some new features that cannot be handled well.
In this study, to accommodate the new UI features (\eg, fragment, navigation, web widget), we fine-tune the UI transition graphs (\sg{}) and propose our definition as follows.

\begin{figure}[!t]
\centering
	\includegraphics[width=.45\textwidth]{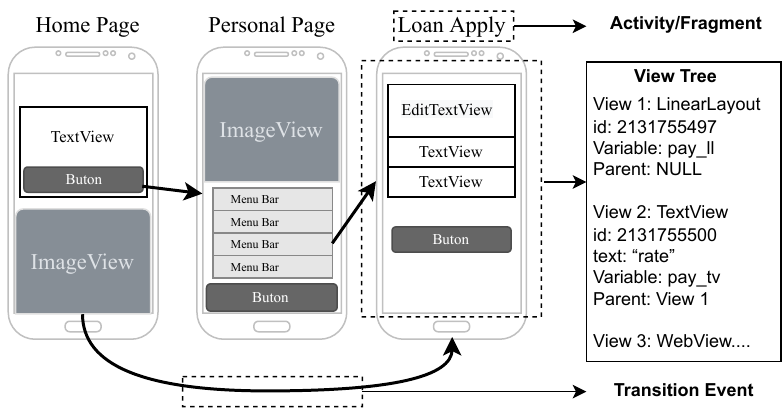}	\caption{Simplified example of a loan app UTG.}
	\label{fig:senario_sample}
    % \vspace{-1em}
\end{figure}

\textit{Definition I:} \tab
A \sg{} is a directed graph $\mathcal{G} = (\mathcal{V},\mathcal{E})$ in a node attribute space $\Omega$, where:
\begin{enumerate}[leftmargin=*]
    \item $\mathcal{V}$ is a node set, we consider 4 categories of windows that users can interact with as a node in a \sg{}: activities, fragments, menus, and dialogs. Activities and fragments are often presented as full-screen windows, while menus and dialogs are short-lived windows that often require the user to take actions.
    \item Each node is assigned GUI view tree attributes in $\Omega$, including native widget attributes (\eg, layout hierarchy, widget type, text), and web widget attributes (\ie, network imprint~\cite{chen2017mass});
    \item Directed edge set $\mathcal{E} \subseteq \mathcal{V} \times \mathcal{V}$ is a set of transitions between activity/fragment and $\varepsilon$ is the single edge within the $\mathcal{E}$.
\end{enumerate}

For better understanding, we show a simplified underground loan app \sg{} in Figure~\ref{fig:senario_sample} as an example. It consists of three activity/fragment nodes with GUI attributes (view tree), and three transition events as the directed edges.

\section{Motivation}
\label{sec:motivation}

In this section, we first illustrate the limitation of the traditional detection approaches. 
As discussed in Section~\ref{sec:intro}, the dynamic approaches are not feasible to perform large-scale detection, hence here we only focus on the static approaches.
After that, we detail the feasibility of our approach by demonstrating the validity of the two key observations.

\subsection{Limitation of the Traditional Approaches}
Features used by those traditional detection approaches can be categorized into the following three categories: \textit{Malicious Payload}, \textit{GUI}, and \textit{Manifest}, as shown in Table~\ref{tab:prewor}.
However, all of these features are ineffective to detect \target{} due to the following reasons:
\begin{itemize}[leftmargin=*]
    \item \textbf{Lack of malicious payload.} 
    %There is no malicious payload in the apps since they are not developed to compromise or damage devices. 
    Most \target{} do not have any malicious payloads (over 80\% through our experiment in Section~\ref{subsec:rq2}), which inevitably leads to the failure of malware detection approaches (\eg, API flow detection). 
    %What's more, the fine-grained (lower than method-level) control flow incurs significant analysis overhead~\cite{frenklach2021android}.
    \item \textbf{Incomplete \& Disguised GUI content.}
    Due to the strict censorship in sensitive areas, most \target{} adopt countermeasures: \textit{(i)} hide the sensitive GUI content (\eg, pictures, text) on remote servers. \textit{(ii)} disguised as \legit{}s, such as using similar icons. 
    These countermeasures result in the failure of content-based detection (see Section~\ref{subsec:rq2}).
    \begin{comment}
     Nowadays, the dynamic loading and web widgets are widely adopted by \target{}. As a result, those approaches~\cite{zhang2014viewdroid,chen2015finding} that only analyze XML widgets become ineffective due to the low identification coverage.
     On the other side, the UI content (\eg, pictures, videos) is always downloaded from the remote server at runtime. Obviously, the detection approaches based on native resources~\cite{nan2015uipicker,li2016peruim,chen2019storydroid} are not feasible due to the high false negative rate. 
     \end{comment}
    \item \textbf{Ineffective Manifest.}
    The Manifest information cannot be used as the unique features to detect \target{}.
    Specifically, the permissions acquired by \target{} are similar to those of \legit{}s~\cite{gao2021demystifying}, 
    while the developer signatures can always be arbitrarily customized (in Android). 
    
\end{itemize}

\subsection{Feasibility of Our Approach}
To perform an effective large-scale detection, we propose an \target{} detection/classification approach based on the \sg{} similarity. Our approach is based on the following two key observations:
\begin{itemize}[leftmargin=*]
\item \textbf{Observation-I:} \sg{}s of \target{} are different from those of the \legit{}s.
\item \textbf{Observation-II:} 
UTGs of the same UEware type are similar, while those of different types vary.
\end{itemize}

For the first observation, we have shown an example of a normal loan app vs. an underground loan app in Figure~\ref{fig:def}. What's more, we additionally select some other apps~\footnote{Due to the page limit, all the details of those examples are shown in \url{https://github.com/HypoopyH/DeUEDroid}.} for comparison. 
We point out that \target{} always do not provide UIs related to terms and conditions and are only served for one purpose (\eg, usury application), which leads to considerable differences in \sg{} structure.
Through our experiment, we even find that the \sg{} of \target{} is different from that of \leg{} ones in statistic (\ie, 11 vs. 22 transition events and 29 vs. 227 widgets, respectively), see details in Section~\ref{subsec:result}.

To demonstrate the second observation, we randomly select another underground financial app (with a different Manifest). 
Although their runtime screenshots are different, they share a similar \sg{} (\eg, home page, loan apply, bank card check). 
And then, we randomly select an underground gambling app for comparison. 
The gambling app provides many game windows rather than the loan applications.  
Due to the different services they provided, the \sg{}s of these two types of apps inevitably vary from each other.

\section{Design Overview}
\label{sec:overview}

\begin{figure}[!t]
\centering
	\includegraphics[width=.45\textwidth]{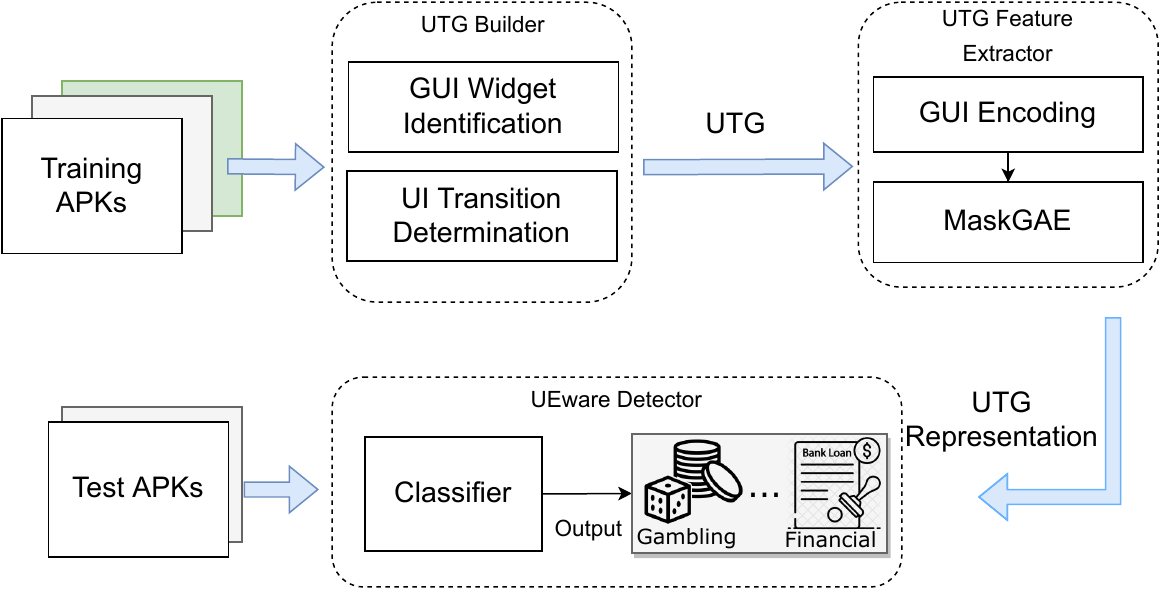}
	\caption{The design overview of \system{}.}
	\label{fig:design_overview}
       \vspace{-1.5em}
\end{figure}

Based on the proposed approach, we design a system named \system{} to detect and classify \target{}. 
Specifically, \system{} first statically builds precise \sg{} for \target{}, and then applies a graph-embedding based method to represent \sg{}.
After that, it leverages the self-supervised method to detect and classify \target{} based on \sg{} similarity.
Figure~\ref{fig:design_overview} shows the architecture of \system{}. There are three modules, \ie, 
\textit{\sg{} Builder}, \textit{\sg{} Feature Extractor} and \textit{\target{} Detector}, as follows:
\begin{itemize}[leftmargin=*]
\item \textbf{\sg{} Builder.}
This module accepts Android APK files as the input, and outputs the corresponding \sg{}s.
It consists of two sub-modules: \textit{GUI widget identification} and \textit{UI transition determination}. Specifically, the first sub-module covers both the native widgets and the web widgets;
while the second sub-module applies a novel UTG construction algorithm to accurately capture UI transitions.
\item \textbf{\sg{} Feature Extractor.}
Based on these \sg{}s, this module combines both UTG topology and GUI attributes together and outputs the \sg{} representation (feature).
Specifically, it first leverages multiple encoding methods to handle GUI attributes of different dimensions. 
After that, it applies a novel algorithm to correlate the graph topology and the GUI attributes.

\item \textbf{\target{} Detector.}
Based on the \sg{} feature, this module leverages the self-supervised learning to train an \target{} classifier, which will be used to perform the detection. Note that the \sg{} based approach allows further classifying the apps into different categories of \target{}. Currently, three underground categories, including \textit{gambling}, \textit{porn} and \textit{financial}, are supported.
\end{itemize}

In the following, we will detail these three modules in Section~\ref{sec:scenario}, Section~\ref{sec:graphai} and Section~\ref{sec:detect}, respectively.

\section{UTG Builder}
\label{sec:scenario}
In this section, we describe the design of \sg{} builder.
This module has two sub-modules: 
\textit{(i)} GUI widget identification, and 
\textit{(ii)} UI transition determination. 

All sub-modules take an Android APK file as the input. 
The output of GUI widget identification is a set of GUI widgets. 
Notably, we collect the layout hierarchy, text, and widget type of native widgets. And the web widget (\eg, WebView) attributes are represented by their network imprints. 
The output of UI transition determination is a directed graph representing the transition relationship between activities/fragments. 
In the end, the \sg{} is well constructed as a directed graph with GUI attributes.

\subsection{GUI Widget Identification}
\label{subsec:gui_identify}
In Android, the GUI widgets can be dived into native widgets and web widgets. Specifically, \system{} proposes different analysis methods for two types of widgets.

\noindent\textbf{Native Widgets.}
As mentioned in Section~\ref{sec:preliminary}, the native widgets can be statically recorded in the XML file or dynamically added in code. To the XML widgets, their widget types (\eg, ImageView, Button), the text (\eg, @string/ic\_hello), and the layout hierarchy (\eg, ConstrainLayout) are well organized in the corresponding XML files. In addition, the activities/fragments load their corresponding XML widgets based on the API calls \textit{findViewbyID}.
To the dynamically added widgets, they are dynamically added in code (\eg, new TextView()),  and their attributes are recorded in the code, which requires taint flow analysis to identify.
These widgets are rendered within the ``\textit{OnCreate}" phase when a activity/fragment is calling up. 

There have been several studies~\cite{xiao2019iconintent,huang2015supor} to perform native widgets analysis. Considering the efficiency, we build up our system based on a SOTA lightweight static UI analysis framework, FrontMatter~\cite{kuznetsov2021all}, to figure out the native widgets.

\noindent\textbf{Web Widgets.}
Nowadays, the hybrid development paradigm is commonly used~\cite{hybridapp}, which acquires remote resources through web widgets.
Specifically, the WebView starts a local browser and invokes the web APIs (\eg,``\textit{LoadURL}") to load a web page.
These widgets are an important part of \sg{}, but are not properly analyzed by native widget analysis studies.

For these web widgets, the essential part is the URL parameter. 
However, web URLs are always dynamically concatenated, which is hard to restore in static analysis~\cite{strings2003,chen2017mass}. 
But we have two observations about web widgets: \textit{(i)} the URL parameter and its related tokens are all in string type.
\textit{(ii)} almost always, a parameter-related token originates from some constant values within the program, such as constants, and XML resources.
To this end, we refer to the previous studies~\cite{chen2017mass} and generate the network imprint to represent the web widget attributes.

Here we use an example (see Figure~\ref{fig:token} in Appendix A) to present the imprint generation. First, the program loads activities/fragments, and searches special API calls (\eg, \textit{WebView.loadURL()}). Once the sink statement is found in Class Y method B, the sink variable is \textit{url}. Then the program will perform backward taint analysis to check whether any variable affects \textit{url}. After running,  \textit{v3} and \textit{X.A()} is determined to affect \textit{url} using \textit{java.lang.StringBuilder.append()}. 
And the \textit{v3} is considered unrelated to any invariant source within the app but instantiated from runtime  variables, so we discard the \textit{v3}. Turn to the \textit{X.A()}, since it is an inter-procedure call, the program additionally loads Class X and sinks the \textit{v1}. Going further, the program finds that \textit{R.string.baseUrl} and \textit{v2}. At this time, \textit{v2} is determined to be an immediate variable, so the program instantiates \textit{v2} with a concrete value and instantiated \textit{R.string.baseUrl} from XML file.
To sum up, \textit{R.string.baseUrl,v2} are instantiated as constant variables, \textit{v3} is instantiated as runtime variable and discarded.  In the end, the program outputs a set of tokens (\eg, [\textit{R.string.baseUrl,v2}]) and removes some common tokens (\eg, \textit{github}). These tokens are used as network imprints to identify web widget attributes.

\begin{algorithm}[!t]
    \footnotesize
	\SetAlgoLined
	\SetKwData{True}{true}
  
	\KwData{$apk$: the APK file.}
	\KwResult{$UTG=(Nodes,Edges)$: the UI transition graph for $apk$.}
	
    \SetKwFunction{FAssociate}{associate}
    \SetKwFunction{Fatgconstruction}{UTG\_Construction}
    \SetKwFunction{Fgetcg}{get\_CallGraph}
    \SetKwFunction{Fgetclass}{get\_AllClasses}
    \SetKwFunction{Fgetmethod}{get\_AllMethods}
    \SetKwFunction{Fgetunits}{get\_AllUnits}
    \SetKwFunction{Fgetcallee}{get\_Callees}
    \SetKwFunction{FgetNavTarget}{get\_NavTargets}
    \SetKwFunction{Fistransition}{is\_Transition}
    \SetKwFunction{Fisfragment}{is\_Fragment}
    \SetKwFunction{Fisinner}{is\_InnerClass}
    \SetKwFunction{Fisact}{is\_Activity}
    \SetKwFunction{FgetfragmentAct}{get\_CallerActs}
    \SetKwFunction{Fhasnav}{has\_Navigation}
    \SetKwFunction{Fgetfragallact}{get\_ActsWithFragment}
    \SetKwFunction{Fgetoutact}{get\_OuterActs}

    \SetKwFunction{FPreprocess}{preprocess}
    \SetKwFunction{FAssociateSamples}{associate\_samples}
    \SetKwFunction{FGetSignature}{GetSignature}
    \SetKwFunction{FGetURLList}{GetURLList}
    \SetKwFunction{FGetSnapshot}{GetSnapshot}
    \SetKwFunction{FAssociateSignature}{AssocSignature}
    \SetKwFunction{FAssociateURLList}{AssocURLList}
    \SetKwFunction{FAssociateSnapshot}{AssocSnapshot}
    \SetKwFunction{Fdoassociate}{do\_associate}
    \SetKwProg{Fn}{Function}{:}{}
    
    \Fn{$\Fatgconstruction{apk}$} {
        $Nodes \gets \{\}$\;
        $Edges \gets \{\}$\;
        $cg \gets \Fgetcg(apk)$\;
        $classes \gets \Fgetclass(apk)$\;
        \For{$class$ in $classes$} {
            \If{$\Fisfragment(class)$ $or$ $\Fisact(class)$}{
                $Nodes~\cup \{class\}$\;
            }
            $methods \gets \Fgetmethod(class)$\;
            \For{$method$ in $methods$} {
                $units \gets \Fgetunits(method,cg)$\;
                \For{$unit$ in $units$}{
                    \If{$\Fistransition(unit)$}{
                         $callees~\cup \Fgetcallee(unit)$\;
                         %\eIf{$\Fisfragment(class)$}{
                            $callers~\cup \FgetfragmentAct(class,cg)$\;
                         %}
                         %{
                        %    $callers~\cup \FgetfragmentAct(class,cg)$\;
                        % }
                         $Edges~\cup \{callers,callees\}$\;
                        }
                }
            }
            \If{$\Fhasnav(class)$}{
                $caller \gets \FgetfragmentAct(class,cg)$\;
                $callees \gets \FgetNavTarget(class)$\;
                $Edges~\cup \{caller,callees\}$\;
            }
        }
        \KwRet $UTG(Nodes, Edges)$\;
    }
    
     \Fn{\FgetfragmentAct{$class,cg$}} {
        \uIf{$\Fisfragment(class)$}{
            $callers \gets \{class,\Fgetfragallact(class,cg)\}$\;
        }
        \uElseIf{$\Fisinner(class)$}{
            $callers \gets \Fgetoutact(class,cg)$\;
        }
        \lElse{
            $callers \gets class$
        }
     \KwRet $\{callers\}$\;
     
    %     \For{$sample_i$ in $samples$} {
    %         $sig$ = \FGetSignature($sample_i$)\;
    %         $url\_list$ = \FGetURLList($sample_i$)\;
    %         $snapshot$ = \FGetSnapshot($sample_i$)\;
    %         $features$[$sample_i$] = $(sig, url\_list, snapshot)$\;
    %     }
    %     \KwRet $features$\;
     }
    
	\caption{The UTG Construction Algorithm}
  \label{alg:atg}
\end{algorithm}

\subsection{UI Transition Determination}
\label{subsec:atg}
The GUI widgets and layout modules are rendered on activities/fragments. And when a new window is opened, it causes a transition. Existing studies~\cite{azim2013targeted,octeau2015composite,chen2019storydroid,yang2018static,liu2022promal}  mainly identify transitions among activities, but overlook the \textit{fragment-related transitions} (\eg, the transitions between fragments and activities) and the \textit{navigation-based transitions} (\eg, navigation among fragments). 
%leading to some omissions. 
As the latter two have been widely adopted to develop Android apps in recent years, the lack of support for them will inevitably lead to imprecise \sg{}. 
To accurately build up the  transition events, we propose a new \textbf{UTG Construction Algorithm}~\ref{alg:atg}.
This algorithm can cover the fragment-related transitions and the navigation-based transitions, and build a precise \sg{}.

Specifically, we first initialize \textit{Nodes} and \textit{Edges} as empty sets, which store the activities/fragments and their transition relationships respectively.
Then we generate the call graph of the given APK, and fetch out all classes defined in the Manifest. 
Note that, there are some implicit run and start pairs (\eg, AsyncTask, OnClickListener, Runnable, and Message) in Android, which need to be bridged to make the call graph complete.

For each class in the APK, we first determine whether it inherits from activity or fragment or not.
If so, it will be added to the \textit{Nodes} set.
Then for all methods in each class, we get all units according to the previous call graph.
For these units, we will determine whether they have transition movements.
Specifically, for the explicitly transition and implicitly transition, they have special APIs (\eg, \textit{StartActivity()}, or \textit{Intent()}). 
By comparing the API signatures, we can locate their transition units.
Besides, the navigation-based transitions are implemented by special structures (\eg, \textit{NavController}). 
We can locate the navigation unit by searching the variable structures.
Finally, there are some hardware events (\eg, \textit{BACK} button), we monitor whether the activity has overloaded system-level calls, to implement the transition.

\begin{figure*}[!t]
\centering
	\includegraphics[width=.95\textwidth]{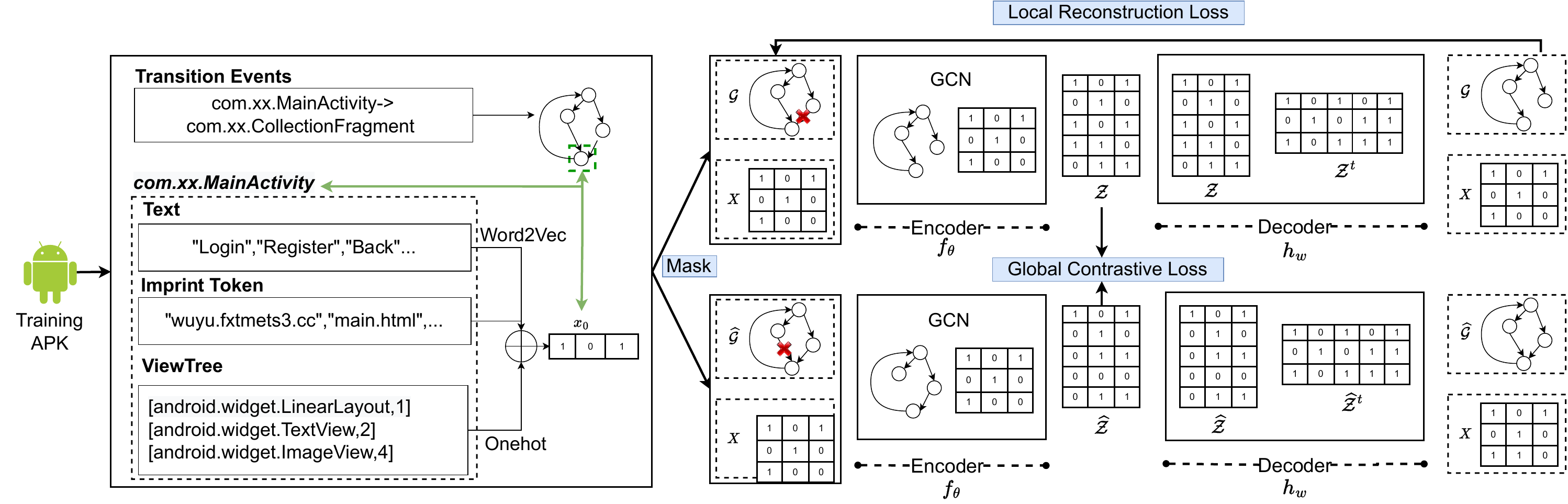}
	\caption{The design of \sg{} feature extractor.}
	\label{fig:graphai}
    % \vspace{-1em}
\end{figure*} 

Furthermore, for each transition unit, we first get the targets (callees) by analyzing the unit arguments based on taint analysis. Due to the difficulty in handling conditional expressions, all possible target activities/fragments are included.
Then we get the source (callers). Specifically, three different processing methods are adopted according to the class type.
\begin{itemize}[leftmargin=*]
    \item If the class is a fragment, the callers are the union of this fragment and all activities that own this fragment. Because activities can trigger transitions of fragments they own. 
    \item If the class is an inner class, the callers are the union of the outer activities/fragments of this inner class.
    \item If the class is not an inner class nor a fragment, the caller is the current class itself. 
\end{itemize}
Finally, for callers and callees found by a transition, we traverse all of them and add all pairs (caller $\to$ callee) to the \textit{Edges} set.

\section{UTG Feature Extractor}
\label{sec:graphai}

In this section, we describe the design of \sg{} feature extractor. The overview of this module is shown in Figure~\ref{fig:graphai}. 
For an input \sg{}, we first encode the GUI attributes $\Omega$ into matrix $X$. And then, we propose a novel algorithm, MaskGAE, to correlate the graph topology $\mathcal{G}$ and GUI attributes $X$ as the \sg{} representation $Z$.

\subsection{GUI Encoding}
To properly integrate different dimension attributes, we adopt multiple encoding approaches.
Specifically, for an activity/fragment node, we first segment the text string and network imprint from GUI widgets. To utilize these string-type attributes, we use a pre-trained Word2Vec~\cite{mikolov2013efficient}  model to generate their representations.
After that, we encode the ViewTree, which consists of multiple GUI widgets. Note that, a GUI widget can be an instance of a system widget class (\eg, button) or a customized widget class inherited from the view class. Thus, we regard the widget type as the GUI widget’s "tag", and use the one-hot model to encode and serialize it. Further, to the layout hierarchy, we traverse the ViewTree and generate a widget sequence.
Finally, we combine all representations into the vector $x_0$ of the node. And since a \sg{} may consist of multiple nodes, the final GUI matrix $X$ is combined by every node representation. Meanwhile, the transition events and the activity/fragment nodes have been represented by the \sg{} topology $\mathcal{G}$.

\subsection{MaskGAE}
In the underground economy, developers always generate new apps by modifying parts of the existing apps.
To counteract the impact of code modifications, it requires high robustness of the graph embedding model. To this end, we propose a novel algorithm, MaskGAE, which adopts the Mask strategy before the \sg{} embedding.
Specifically, the Mask strategy can be viewed as an adversarial attack that provides a new graph as data augmentation.   
In this paper, we adopt Edge-wise random masking, which is defined as:
\begin{equation}
    \varepsilon_{mask} \sim Bernoulli(p)
\end{equation}
where $\varepsilon$ denotes the edge within the $\mathcal{E}$.

After the mask, we follow the success of the graph autoencoder (GAE)~\cite{hinton1993autoencoders}, which is designed to reconstruct graph inputs, to correlate the graph topology and GUI attributes. 
In this study, our encoder $f_{\theta}$ is 2-layer graph convolutional networks (GCN)~\cite{chiang2019cluster}, a well-established GNN architecture. 

Notably, due to the masking strategy, our loss calculation consists of two parts: local reconstruction loss $L_{local}$ and global contrastive loss $L_{global}$.
For the masked graph, we divide the edges as remaining edges and masked edges. The masked edges are selected as positive samples, while the disconnected node's edges are selected as negative samples. For the embeddings from the decoder, we calculate their inner product as the probability of the edge existing. 
\begin{equation}
\begin{aligned}
    \mathcal{L}_{Local} = - ( \frac{1}{|\varepsilon^+|} \sum_{(u,v) \in \varepsilon^+} \log{h_w(z_u,z_v)}  \\ + \frac{1}{|\varepsilon^-|} \sum_{(u^{'},v^{'}) \in \varepsilon^-} \log{(1-h_w(z_{u^{'}},z_{v^{'}}))}
    )
\end{aligned}
\end{equation}
where $\mathcal{Z}$ is the graph embedding result from the encoder; $\varepsilon^+$ is a set of positive edges while $\varepsilon^-$ is a set of negative edges sampled from the graph. 

Further, we calculate the distance between the two representations $(Z, \widehat{Z})$ as the global contrastive loss. Specifically, we normalize the loss value into the range of $[0, 1]$ to facilitate optimization.

\begin{equation}
    \mathcal{L}_{Global} = \frac{\sum_{i=1}^N (z_i-\widehat{z}_i)^2}{N}
\end{equation}
Finally, we combine the global loss and local loss into the total loss and use gradient descent to minimize it.

\begin{equation}
    \mathcal{L}_{Total} = \mathcal{L}_{Local} + \alpha\mathcal{L}_{Global}
\end{equation}
Where $\alpha$ denotes a  hyperparameter trading off two terms.
\section{\target{} Detector}
\label{sec:detect}
\begin{figure}[!t]
\centering
	\includegraphics[width=0.45\textwidth]{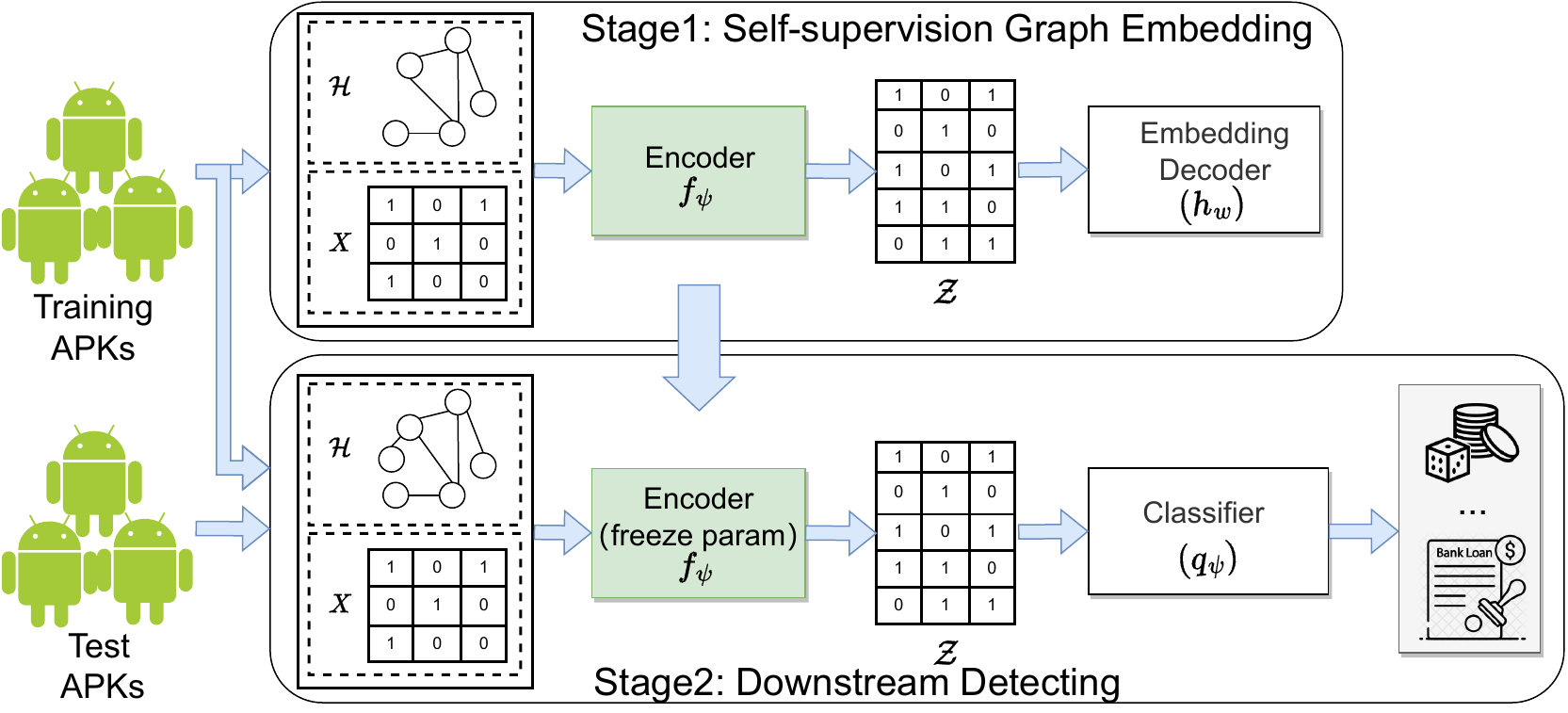}
	\caption{The design of \target{} detector.}   
        \vspace{-1.5em}
	\label{fig:detecting}
\end{figure}

In this section, we describe the design of \target{} detector. 
The \target{} detector leverages the self-supervised learning, which is widely adopted in previous studies~\cite{velickovic2019deep,hassani2020contrastive}. 
In detail, it consists of two stages: self-supervision training task, and downstream training task, see Figure~\ref{fig:detecting}.

The self-supervised training task also refers to GAE, however, it is different from the \sg{} feature extractor.
In detail, the self-supervision training task accepts the APK relation graph $\mathcal{H}$ and the APK feature $X$, and trains the encoder $f_\psi$.
Referring to previous studies~\cite{zhang2018name}, the APK relation graph $\mathcal{H}$ is an undirected graph, while the node of the graph is APK, and the edge of the graph is the overlap Manifest information (\ie, PackageName, AppName, developer signature) between APKs. Meanwhile, the APK feature $X$ is the \sg{} representation produced by encoder $f_\theta$.
And the downstream task uses the encoder $f_\psi$ to train the classifier $q_\psi$ for \target{} detection.

\noindent \textbf{Classifier:}
The APK encoder $f_\psi$ learned from self-supervision training is set frozen and directly introduced into the downstream task. 
According to this frozen encoder, the downstream task then calculates the $\mathcal{Z}$. 
Finally, the classifier $q_\psi$ (also called the downstream decoder) is trained based on the labeled dataset. In this study, we leverage a linear classifier (\ie, a logistic regression model) and the labeled information is the app types (\ie, underground gambling, underground porn, underground financial, and \legit{}s).
Specifically, the detection task decoder parameter is calculated as:
\begin{equation}
    \mathcal{\psi} = argmin\mathcal{L}_{sup}(f_{\theta},q_{\psi},\mathcal{H},y)
\end{equation}
where $\mathcal{H}$ is the APK relation graph and y is the label of \target{}.

In the end, when input test APKs, the \system{} first combine the test APKs and training APKs together to construct a new graph $\mathcal{H}$. And then, the \system{} labels the test APKs through the encoder $f_\psi$ and decoder $q_\psi$. 

\section{Implementation and Evaluation}
\label{sec:evaluation}
We have implemented a prototype of \system{}. Specifically, we leverage Soot~\cite{vallee2000optimizing} and FlowDroid~\cite{arzt2014flowdroid} for taint analysis and Frontmatter~\cite{kuznetsov2021all} for GUI analysis. 
In network imprint analysis, we additionally adopt Java string analysis techniques~\cite{strings2003}, and customize the conditions of instantiate.
We deploy \system{} on a server has 16 cores with 2.1GHz CPU, 256GB memory, and 8TB hard drives. 

Furthermore, we will evaluate \system{} by answering the following three research questions:
\begin{itemize} [leftmargin=*]
    \item \textbf{RQ1}: How effective is \system{} in \sg{} construction?
    \item \textbf{RQ2}: How effective is \system{} in \target{} detection? 
    \item \textbf{RQ3}: How effective and efficient is \system{} in large-scale \target{} detection? 

\end{itemize}

\subsection{Evaluation Setup}
\label{subsec:setup}

For RQ1, we investigate the capability of \sg{} construction. To evaluate the capability, we build up the \textbf{source-available dataset}, consisting of \num{5} self-developed apps and \num{10} real-world apps.
Specifically, we developed \num{5} apps as our ground-truth benchmark, which covers different transitions (\ie, the transitions between activity - activity, activity - fragment, fragment - fragment, and navigation-based transitions). 
This method is used to accurately understand the ground truth of transitions and widgets, and it has been widely adopted by previous studies~\cite{chen2019storydroid,li2014know}. 

In addition, to make our experiments more convincing, we compare the SOTA works (\ie, IC3~\cite{octeau2015composite}, Gator~\cite{rountev2014static}) and analyze several famous open-sourced apps~\footnote{The package names are: \textit{org.woheller69.wather, site.leos.setter, ademar.bitac, de.digisocken.antherrss, net.gsantner.dandelior, com.github.dfa.diaspora\_android, com.chao.app, com.kylecorry.trail\_sense, org.woheller69.arity, de...osmplugin} (this name was truncated due to its excessive length).}. in f-droid~\cite{F-droid}. 
Note that there are some studies~\cite{chen2019storydroid,chen2017mass} that focus on static transition identification or imprint generation. However, we cannot compare them since they do not release the code or dataset.

\begin{table}[!tb]
\caption{The dataset for evaluation.}
\centering
\small
\resizebox{.45\textwidth}{!}{
\begin{tabular}{c|c|c|l|c} 
\toprule
%\hline
Dataset      & \multicolumn{1}{c|}{ App Type} & Setup Time     & \multicolumn{1}{c|}{Detail} & Number  \\ 
\midrule
\multirow{3}{*}{\begin{tabular}[c]{@{}c@{}}Source-available\\ 
Dataset\end{tabular}} & Self-developed   & July, 2022   & Self-developed Apps   & 5 \\ 
%\cline{4-5}
% &   &   & Renaming Obfuscation Apps   & 5 \\ 
%\cline{4-5}
% &   &   & Code Obfuscation Apps & 5 \\ 
%\cline{4-5}
% &   &   & Reflect Obfuscation Apps & 5 \\ 
\cline{2-5}
& Real-world   & 2021-2022   & Real-world Apps   & 10 \\ 
\cline{2-5}
% & \multirow{6}{*}{Real-World} & \multirow{6}{*}{2021-2022}   & site.leos.setter    & 1 \\ 
%\cline{4-5}
% &   &   & de.digisocken.antherrss    & 1 \\ 
%\cline{4-5}
% &   &   & org.woheller69.weather    & 1 \\
% \cline{4-5}
% &   &   & net.gsantner.dandelior    & 1 \\ 
% \cline{4-5}
% &   &   & Real-world Apps    & 1 \\ 

%\cline{2-5}
 & Total   & ~ & ~  & 15      \\ 

\midrule
\multirow{6}{*}{\begin{tabular}[c]{@{}c@{}}Ground-truth \\ Dataset\end{tabular}} & \multirow{3}{*}{\target{}}    & \multirow{3}{*}{June, 2022} & Gambling & \num{310} \\ 
\cline{4-5}
 &   &   & Porn     & \num{441} \\ 
\cline{4-5}
 &   &   & Financial    & \num{97} \\ 
%\cline{4-5}
% &   &   & Miscellaneous      & \num{453} \\ 

\cline{2-5}
 & Normal Apps    & June, 2022 & -  & \num{852}    \\ 
\cline{2-5}
 & Total   & ~ & ~  & $1,700$    \\ 
\midrule
\multirow{3}{*}{\begin{tabular}[c]{@{}c@{}}Large-scale \\ Dataset\end{tabular}}  & Wild     & June-July, 2022 & Apps from websites    & $13,460$  \\ 
\cline{2-5}
 & Appstore   & June-July, 2022 & Apps from AppStore     & $10,557$   \\ 
\cline{2-5}
 & Total   & ~ & ~  & $24,017$   \\
\bottomrule
\end{tabular}
}

\label{tab:datasets}
% \vspace{-2.5em}
\end{table}

For RQ2, we evaluate the capability in \target{} detection and classification. 
As there does not exist any publicly available dataset. 
we have to make efforts to build the \textbf{ground-truth dataset} to perform the evaluation, as follows:
\begin{itemize}[leftmargin=*]
    \item  First, five experienced experts in our team collect wild apps from social media, forums, and websites. They work independently to download, install, and launch these apps, including registering accounts and investigating their usage processes.
    \item Second, these experts select apps that provide sensitive services (\ie, Porn, Gambling, Financial in this work) for further analysis. According to different regulations in special fields (such as asset review in loan, etc.), they independently check whether the app usage processes comply with the required behavior patterns, and mark the non-compliant apps as UEware.
    \item Finally, all experts work together to make the final decision by majority voting. The ground-truth dataset consists of $1,700$ apps, including \num{310} gambling apps, \num{441} porn apps, \num{97} financial apps, and \num{852} \legit{}s.
\end{itemize}

We randomly select \num{70}\% of them as the training set, \num{20}\% as the validation set, and \num{10}\% as the test set.

For RQ3, we conduct a large-scale experiment to evaluate the performance of \system{} to mitigate real-world threats.
For this goal, we collect a large number of real-world apps and setup the \textbf{large-scale dataset}. Based on the source of acquisition, we divide these apps into two categories: AppStore apps and wild apps.
The AppStore apps are collected from formal app markets, such as Mi Store, 360 Software Manager. In contrast, wild apps are collected from informal channels, such as social platforms or websites.
In total, the large-scale dataset consists of $24,017$ apps ($10,557$ AppStore apps and $13,460$ wild apps).
Based on this dataset, we evaluate the performance (including the detection rate and the time consumption), and we also analyze the detection results.

In total, we setup three datasets (\ie, source-available dataset, ground-truth dataset, and large-scale dataset, as shown in Table~\ref{tab:datasets}) to perform the evaluation. Note that all hardened apps that cannot be analyzed have been removed from our datasets.

\subsection{Evaluation Result}
\label{subsec:result}
\subsubsection{For RQ1}

App sketch construction has two sub-modules: \textit{(i)} GUI widget identification, and \textit{(ii)} UI transition determination.

First, for the GUI widget identification, \system{} applies Frontmatter~\cite{kuznetsov2021all}, a SOTA GUI analysis tool to identify the native widgets. 
So here we only evaluate the effectiveness of web widget identification, which is implemented by ourselves. And we list our evaluation result in Table~\ref{tab:static_eva_imprint}.
For the self-developed apps, the token number is clearly identified by our developers. In total, the self-developed apps have \num{36} tokens, such as IP host, and domain name.
In the end, \system{} identifies \num{36} tokens from the self-developed apps, with  \num{3} false positive and \num{8} false negative (F1-score is \num{84.7}\%). 
For real-world apps, we first manually identify the tokens used by their web widgets. Specifically, we insert logging code (\ie, logging.log()) in the source code after the network API, which does not affect the app function. Then we run the app to generate the log, and also inspect the source code to get the final tokens.
The F1-score of real-world apps are range from \num{75.0}\% to \num{95.5}\%, which shows that our imprint generation has sufficient coverage.

\begin{table}[!tb]
\caption{The evaluation of imprint generation.}
\centering
\resizebox{.45\textwidth}{!}{
\begin{tabular}{lccl}
\toprule
\textbf{Dataset Component}   & Token (\#)  &  Identified (F1-score) & \\
\midrule
\textbf{Self-Developed App}   & 36 & 84.7\%  \\
%$\qquad\vdash$\# No Obfuscation & 36 & 84.3\%  \\
%$\qquad\vdash$\# Rename Obfuscation  & 36 & 84.3\%  \\
%$\qquad\vdash$\# Code Obfuscation  & 36 & 84.3\% \\
%$\qquad\vdash$\# Reflect Obfuscation  & 36 & 84.3\%  \\
%\midrule
\textbf{Real-World App}  &  &   \\
$\qquad\vdash$\# site.leos.setter & 21 & 95.5\%\\
$\qquad\vdash$\# de.digisocken.anotherrss & 40 & 90.9\% \\
$\qquad\vdash$\# org.woheller69.weather & 43 & 90.5\% \\
$\qquad\vdash$\# net.gsantner.dandelior & 10 & 88.9\% \\
$\qquad\vdash$\# com.github.dfa.diaspora\_android & 12 & 80.0\% \\
$\qquad\vdash$\# com.chao.app & 50 & 83.3\% \\
$\qquad\vdash$\# de.storchp.opentracks.osmplugin & 47 & 75.0\% \\
$\qquad\vdash$\# com.kylecorry.trail\_sense & 9 & 84.2\% \\
$\qquad\vdash$\# org.woheller69.arity & 19 & 90\% \\
$\qquad\vdash$\# ademar.bitac & 8 & 85.7\% \\
\midrule
\textbf{Total}  & 295 & 85.4\%  \\

\bottomrule
\end{tabular}
}
\label{tab:static_eva_imprint}
% \vspace{-2em}
\end{table}

\begin{table}[!tb]
\caption{The evaluation of UI transition determination. The \textbf{boldfaced} score
denotes the best result.}
\centering
\resizebox{.45\textwidth}{!}{
\begin{threeparttable}
\begin{tabular}{llllllllll}
\toprule
\multirow{2}{*}{\textbf{Dataset Component}}   & \multicolumn{2}{c}{Transition}    &  \multicolumn{3}{c}{Identified (F1-score)} & \\
\cmidrule{2-3}\cmidrule{4-6}
 & Type & \#  & Gator & IC3 & \system{} \\
\midrule
\textbf{Self-Developed App}  &  All & 62 & 17.6\% & 57.4\% & \textbf{97.6\%}     \\
%$\qquad\vdash$\# No Obfuscation       &  All & 62 & 57.4\% & \textbf{97.6\%} \\
$\qquad\vdash$\# App1   & Act-Act\tnote{1}        & 13 & 26.7\%  & 96.0\%  & \textbf{96.3\%} \\
$\qquad\vdash$\# App2   & Act-Frag.        & 13 & 26.7\%  & 76.2\%  & \textbf{92.3\%}   \\
$\qquad\vdash$\# App3   & Frag.-Frag.   & 13 & 0\%   & 0\%  & \textbf{100\%}  \\
$\qquad\vdash$\# App4   & Navigation  & 12 & 0\% & 0\% & \textbf{100\%} \\
$\qquad\vdash$\# App5 & All  & 11    & 30.8\%  & 62.4\% & \textbf{100\%} \\
%$\qquad\vdash$\# Rename Obfuscation  & All & 62  & 57.4\% & \textbf{97.6\%} \\
%$\qquad\vdash$\# Code Obfuscation  & All & 62  & 57.4\% & \textbf{97.6\%} \\
%$\qquad\vdash$\# Reflect Obfuscation  & All & 62 & 57.4\% & \textbf{97.6\%} \\
%$\qquad\vdash$\# String Encryption  & - & 62  & 97.6\% & 57.4\% \\
\midrule
\textbf{Real-World App}  &  &      &       &   &  \\
$\qquad\vdash$\# site.leos.setter & All & 5 & -\tnote2{}  & 0\%  & \textbf{72.7\%} \\
$\qquad\vdash$\# de.digisocken.anotherrss & All & 4 & 42.9\%  & 50\% & \textbf{80.0\%} \\
$\qquad\vdash$\# org.woheller69.weather & All & 12 & - & - & \textbf{91.7\%}  \\
$\qquad\vdash$\# net.gsantner.dandelior & All & 4 & 50.0\% & 57.2\% & \textbf{75.0\%}  \\
$\qquad\vdash$\# com.github.dfa.diaspora\_android & All & 5 & 47.0\% & 50.0\% & \textbf{88.9\%}  \\
$\qquad\vdash$\# com.chao.app & All & 12 & - & - & \textbf{96.0\%}  \\
$\qquad\vdash$\# de.storchp.opentracks.osmplugin & All & 10 & - & 53.3\% & \textbf{90\%}  \\
$\qquad\vdash$\# com.kylecorry.trail\_sense & All & 6 & 33.3\% & - & \textbf{83.3\%}  \\
$\qquad\vdash$\# org.woheller69.arity & All & 5 & 52.8\% & 88.9\% & \textbf{88.9\%} \\
$\qquad\vdash$\# ademar.bitac & All & 9 & - & 66.7\% & \textbf{90\%}  \\
\midrule
\textbf{Total}  & All &  134  &  24.6\%  & 57.0\%  & 
\textbf{92.3}\%  \\
\midrule
\end{tabular}
\begin{tablenotes}
\small
%\item[1] https://f-droid.org/en/packages/site.leos.setter/
%\item[2] https://f-droid.org/en/packages/de.digisocken.anotherrss/
%\item[3] https://f-droid.org/en/packages/org.woheller69.weather/
\item[1] Act. means Activity, and Frag. means Fragment.
\item[2] timed out within the 30 minute time limit.
\end{tablenotes}
\end{threeparttable}
}
\label{tab:static_eva_atg}
% \vspace{-1em}
\end{table}

Second, for the UI transition determination, we list our evaluation result in Table~\ref{tab:static_eva_atg}.
Since the three components (\ie, dialog, menu, activity) have been  well studied in previous studies~\cite{liu2022promal,kuznetsov2021all,octeau2015composite}, we mainly focus on the new android development features (\ie, fragment and navigation) in this work.
Specifically, these self-developed apps are designed to be developed by certain transition types.
The \textbf{App1} to \textbf{App4} consist of a single transition type (\eg, activity-fragment) to evaluate the accuracy of different transition types, and \textbf{App5} is implemented by all transition types.
From the evaluation results, it can be seen that the accuracy of Gator is low. This is because Gator does not support advanced SDK and does not account for fragment-related transitions and navigation-based transitions.
Besides, IC3 can accurately identify the transition between Activity, but also have no ability to identify the fragment-related transitions.
What's more, to the real-world apps, the Gator and IC3 are also ineffective (even the highest F1-score lower than 88.9\%) and even can not get the result within \num{30} minutes.
In contrast, \system{} can identify all transition types (especially fragment-related transitions). To real-world apps, \system{} is able to produce results within the specified time, and also receives high accuracy and even the lowest F1-score is more than \num{72.7}\%.
In total, \system{} achieves high F1-score in transition identification (92.3\%), which has significantly improvement  from \num{35.5}\% to \num{67.7}\% than previous studies.

We further explain the evaluation results. First, since we take into account new transition types (\ie, fragment-activity transition, and navigation-based transitions) besides activity transition, \system{} can label all kinds of sensitive APIs and perform a more accurate algorithm to identify the transition pairs. It results in the high accuracy of UI transition determination, especially the fragment-related transition pairs.
Second, since we set a maximum taint depth to ensure time efficiency, it leads to some omissions of transition identification. For example, we show a missing edge in the real-world app, the missing edge consists of a long call chain (\ie, \textit{onClick() - func1() - func2() - ... - func9() - actionStart() - startActivity()}) that beyond our maximum recursion.

\begin{table*}[!tb]
\caption{The \target{} detection F1-score (\%) and \target{} category classification accuracy (\%) of different features and algorithms. In each column, the \textbf{boldfaced} score denotes the best result.}
\centering
\resizebox{.9\textwidth}{!}{
% \usepackage{multirow}
\begin{tabular}{llcccc}
\hline
\multirow{2}{*}{}     & \multirow{2}{*}{Algorithm}        & \multicolumn{2}{c}{Manifest}     & \multicolumn{2}{c}{UTG}   \\ \cline{3-4} \cline{5-6}
  &    & \multicolumn{1}{l}{Detection F1-score} & \multicolumn{1}{l}{ Classification Accuracy} & \multicolumn{1}{l}{ Detection F1-score} & \multicolumn{1}{l}{Classification Accuracy} \\ \hline
  \multirow{4}{*}{Supervised}      &
  LR    & $76.81_{\pm 0.33}$ & $79.41_{\pm 0.92}$   & $91.00_{\pm 0.45}$ & $88.65_{\pm 0.76}$   \\
  & GAT   & $79.40_{\pm 0.43}$ & $82.56_{\pm 0.33}$   & $92.17_{\pm 0.05}$ & $93.64_{\pm 1.75}$   \\
  & GraphSAGE      & $91.00_{\pm 0.00}$ & $81.18_{\pm 0.32}$    & $97.83_{\pm 0.12}$ & $93.66_{\pm 1.10}$   \\
  & GCN   & $89.24_{\pm 0.37}$ & $81.89_{\pm 0.37}$   & $96.76_{\pm 0.24}$ & $93.73_{\pm 1.36}$   \\ \hline
\
    \multirow{2}{*}{Self-supervised} & GAE & $92.29_{\pm 0.18}$      & $80.76_{\pm 1.14}$  & $97.94_{\pm 0.07}$      & $98.38_{\pm 0.41}$  \\
  & MaskGAE (Ours) & $\textbf{92.54}_{\pm 0.21}$        & $\textbf{82.61}_{\pm 1.52}$ & $\textbf{98.22}_{\pm 0.06}$        & $\textbf{98.97}_{\pm 0.22}$ \\ 
  \bottomrule
\end{tabular}
}
%\begin{tablenotes}
%\footnotesize
%\item[1] Manifest consists of the basic app information (\ie, app name, package name and certificate hash).
%\end{tablenotes}

\label{tab:ai_result}
%\vspace{-1em}
\end{table*}

\begin{tcolorbox}[size=title,colback=white]
\textbf{Answer RQ1:}
\system{} is effective in \sg{} construction that covers new UI features. Specifically, \system{} can accurately identify imprint tokens (over \num{85}\% F1-score) and transition events (achieves 92.3\% F1-score, a far better performance than the SOTA systems).
\end{tcolorbox}

\subsubsection{For RQ2}
\label{subsec:rq2}

To answer RQ2, we evaluate the effectiveness of our system in \target{} detection and classification.
Specifically, we first evaluate the traditional signature-based detection methods (\ie, malware detection, GUI content detection) on \target{}.
And then, we evaluate \system{} and select five SOTA  algorithms (\ie, LR, GAT~\cite{velivckovic2017graph}, GraphSAGE, GCN~\cite{chiang2019cluster}, GAE) as the baseline, and further compare with other SOTA machine learning based detection methods (\ie, Drebin~\cite{arp2014drebin}, MamaDroid~\cite{mariconti2016mamadroid}).

First, we evaluate the effectiveness of the signature-based detection methods.
For malware detection methods, we randomly select \num{200} \target{} and upload them to VirusTotal, the largest online malware detection platform with 63 security vendors, containing the vast majority of malicious code signatures. However, we find that only \num{34} apps have been labeled as malware, while the rest are considered benign or caused only one vendor warning. This phenomenon proves that \textit{ the malware detection methods are inefficient for \target{} detection and have a high false negative rate (over \num{80}\%).}
And then, we evaluate the GUI content detection methods. After decoding the APKs, we parse their local figures/texts and perform the manual review. Apart from the general icons (\eg, button icon, and payment bank icon), there is almost no sensitive content (\eg, porn picture, gambling picture) in local APKs (over 95\%). 
To this end, we can only leverage their icons to perform the detection. Specifically, we use the seresnext50 network, a widely used network in the previous image learning studies~\cite{mahbod2020transfer,park2019study}, to encode the icon and perform the detection. However, the F1-score in \target{} detection is only 45.7\%.
After manual review, the main reason is that many \target{} imitate the icons of \leg{} ones, resulting in a high false positive rate of content-based detection.
\textit{The incomplete \& disguised GUI content results in the failure of the content-based detection methods. }
The above evaluations justify our motivation in Section~\ref{sec:motivation}.

\begin{figure}[!t]
\centering
	\vspace{2mm}\includegraphics[width=0.45\textwidth]{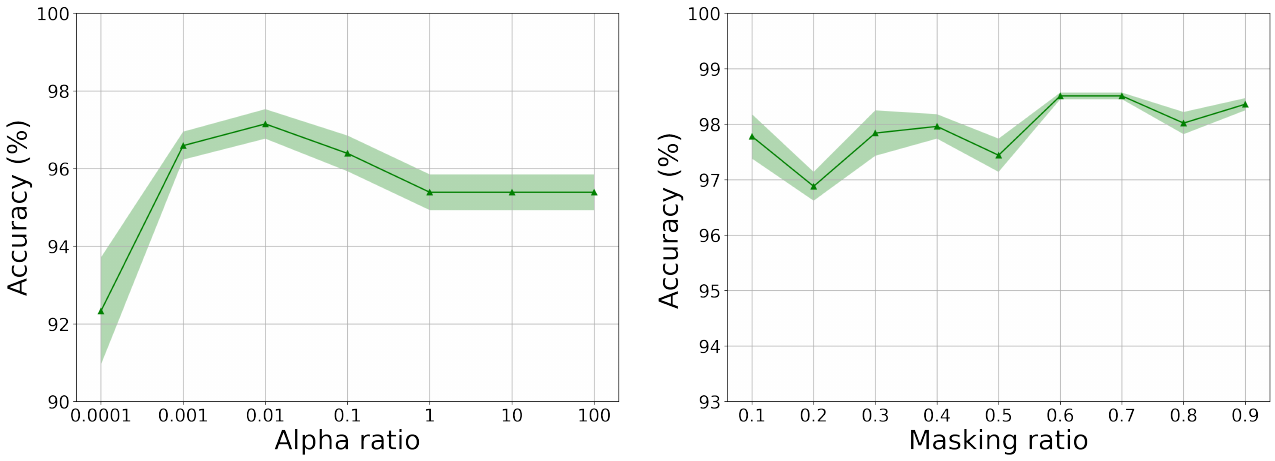}
	\caption{The effect of Alpha and mask ratio to \system{}}
	\label{fig:ratio}
    % \vspace{-1.5em}
\end{figure}

Second, we evaluate our system capability on the ground-truth dataset. Our evaluation is based on two metrics: the detection capability and the classification capability.
The detection capability can be regarded as a binary classification problem. In our dataset, the number of normal apps is roughly equal to the number of UEware. We use F1-score to measure the detection performance.
On the other hand, the classification capability is a multi-class classification problem. Since the distribution of the dataset may not be even, we use classification accuracy to measure the performance.

Specifically, we use the Manifest information (\ie, permission, app size, package name, app components), and five SOTA algorithms as the baseline. 
We closely follow the linear evaluation scheme as previous studies~\cite{velickovic2019deep} and report the detection F1-score and classification accuracy in Table~\ref{tab:ai_result}.

\begin{table}[!tb]
\caption{The comparison with SOTA AI-based malware detection methods. The results contain the detection F1-score and classification accuracy.}
\centering
\resizebox{.45\textwidth}{!}{
% \usepackage{multirow}
\begin{threeparttable}
\begin{tabular}{lllcc}
\hline
\multirow{2}{*}{Method}   & \multirow{2}{*}{Feature}     
& \multirow{2}{*}{Algorithm}  & \multicolumn{2}{c}{Result}    \\ \cline{4-5} &  &
  & \multicolumn{1}{l}{F1-score} & \multicolumn{1}{l}{  Accuracy}  \\ \hline
Drebin & Manifest \& API~\tnote1{}  & DNN & 84.59\% & 79.00\% \\ 
MamdDroid & API Markov Chain  & RF~\tnote2{} & 96.94\% & 98.32\% \\
DeUEDroid & \sg{}  & MaskGAE & 98.22\% & 98.97\% \\
  
  \bottomrule
\end{tabular}
\begin{tablenotes}
\small
%\item[1] https://f-droid.org/en/packages/site.leos.setter/
%\item[2] https://f-droid.org/en/packages/de.digisocken.anotherrss/
%\item[3] https://f-droid.org/en/packages/org.woheller69.weather/
\item[1] Drebin uses the Manifest feature and some chosen APIs as composite features.
\item[2] Random Forest.
\end{tablenotes}
\end{threeparttable}
}
%\begin{tablenotes}
%\footnotesize
%\item[1] Manifest consists of the basic app information (\ie, app name, package name and certificate hash).
%\end{tablenotes}

\label{tab:aiCompare}
% \vspace{-2em}
\end{table}

\begin{itemize}[leftmargin=*]
    \item First, we compare the evaluation result between \sg{} (column\#5-6) and baseline (column\#3-4).  The result shows that the \sg{} feature can improve the detection performance of all algorithms, with outstanding F1-score/accuracy improvement than Manifest across all algorithms (ranging from \num{5.68}\% to \num{14.19}\% in \target{} detection and \num{9.24}\% to \num{17.62}\% in \target{} classification). The outstanding improvement of the comparative experiments shows the \sg{} feature is universal for all algorithms and is effective in \target{} detection and category classification.
    \item Second, we compare the evaluation result of the \system{} algorithm, MaskGAE (line\#7) with other SOTA algorithms (line\#2-6).  And our algorithm achieves leading performance among all algorithms, which exceeds all the supervised algorithms in \target{} detection and classification, and even exceeds GAE \num{0.28}\% and \num{0.59}\% in F1-score and classification accuracy.  This proves that our algorithm is effective, can better handle complex unstructured GUI attributes, and performs well on topological-attribute mixed information. Further, we show the Alpha and mask ratio effect of our detection results in Figure~\ref{fig:ratio}. From the alpha ratio effect in Figure~\ref{fig:ratio}, we can see that by increasing the alpha ratio from \num{0.0001} to \num{100}, the accuracy first smoothly improves to \num{97.15}\% and declines then to \num{95.39}\%. The gap of ratio influence fluctuates very little, only around \num{4.82}\%. And the same as the masking ratio, the ratio influence gap is around \num{0.73}\%.  The ratio effect indicates that our algorithm is very stable and insensitive to the influence of hyper-parameters, which can guarantee stable detection results.
\end{itemize}

In the end, \system{} achieves \num{98.22}\% F1-score in \target{} detection and \num{98.97}\% accuracy rate in \target{} classification.
To demonstrate the effectiveness, we also compare some SOTA machine learning based detection methods. 
Specifically, we compare our approach with two representative works, i.e., Drebin~\cite{arp2014drebin} and MamaDroid~\cite{mariconti2016mamadroid}~\footnote{Drebin and MamaDroid are implemented according to the previous studies~\cite{zhang2020enhancing}, and we will also release the implementation on our github~\cite{DeUEDroidtool}.}.
Table~\ref{tab:aiCompare} gives the result. Obviously, our approach achieves better performance in both binary detection and multi-class classification tasks.

Specifically, Drebin relies on features such as Manifest and sensitive API, which makes the UEware detection ineffective due to the essential limitation (see Section~\ref{sec:motivation}). 
While MamaDroid uses the global call graph to perform the detection. The number of API calls within the graph grows over time~\cite{mariconti2016mamadroid}, which makes the entire API graphs of apps become a heavy-weight feature. As a result, MamaDroid's model would be extensively disturbed to affect the time consumption (more than half of the apps in our dataset take over 10 minutes).

\begin{tcolorbox}[size=title,colback=white]
\textbf{Answer RQ2:}
\system{} demonstrates effectiveness in \target{} detection and classification, achieving a 98.22\% detection F1-score and a 98.97\% classification accuracy, outperforming traditional approaches. Besides, its algorithm surpasses SOTA algorithms and exhibits insensitivity to hyper-parameters.
\end{tcolorbox}

\subsubsection{For RQ3}
\label{subsubsec:rq3}
We apply \system{} on a large-scale dataset
to evaluate its real-world performance, focusing on detection results, time efficiency, and \sg{} measurement.

\begin{figure}[!htb]
\centering
	\includegraphics[width=0.45\textwidth]{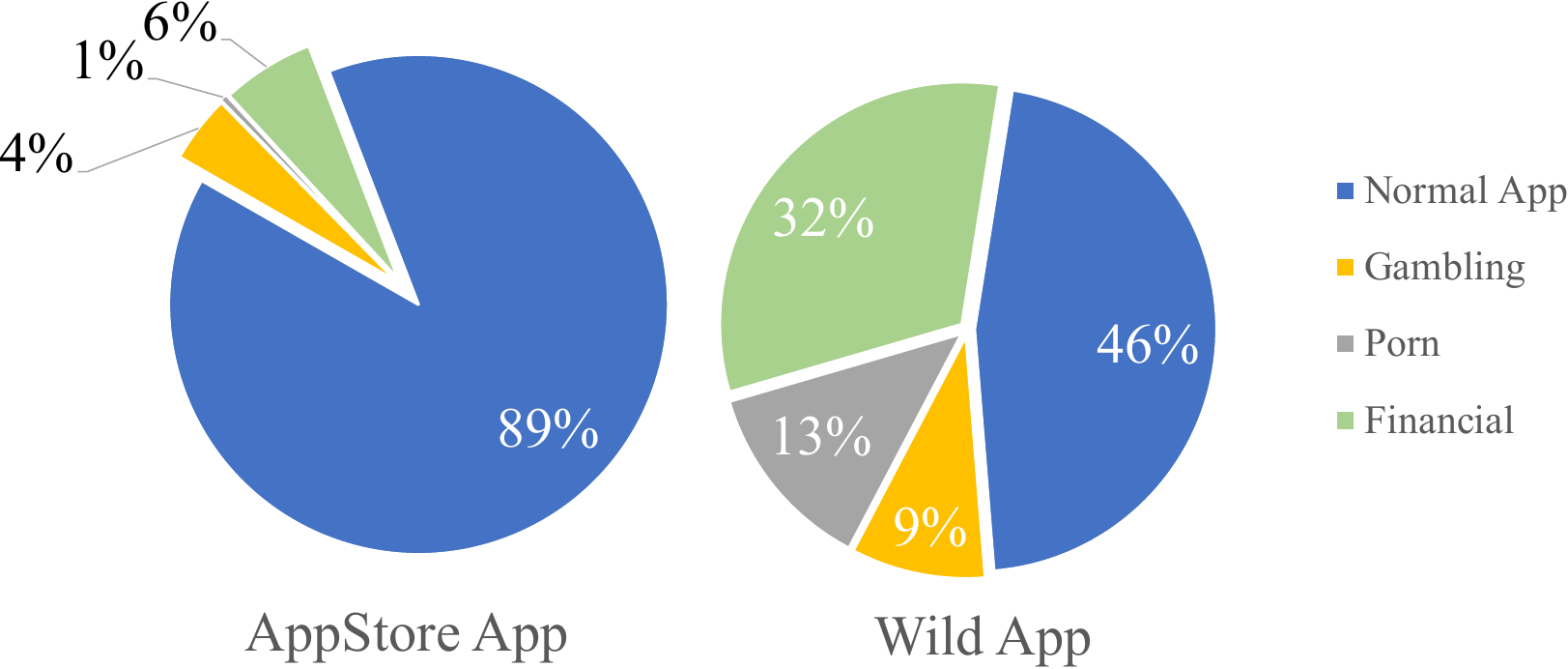}
	\caption{Detection results on the large-scale dataset.}
	\label{fig:performance}
    % \vspace{-1.5em}
\end{figure}

\noindent\textbf{Large-scale Detection Result.}
The detection results on the large-scale dataset are remarkable, as illustrated in Figure~\ref{fig:performance}. First, we analyze the detection results of wild apps, finding that \textbf{more than half (54\%) of the wild apps are \target{}} (\num{9}\% gambling, \num{13}\% porn, and \num{32}\% financial), while the remaining \num{46}\% are \leg{}s. Next, we examine the detection results for AppStore apps and surprisingly discover that \textbf{11\% of AppStore apps are identified as \target{}} (\num{4}\% gambling, \num{1}\% porn, and \num{6}\% financial).

To further confirm the detection result, we additionally select detected \target{} from the two dataset components for manual review.
In total, we review a total of \num{400} apps, and  \num{356} apps are consistent with the manual review results. Since the Out-of-distribution (OOD) of the dataset, the accuracy in the large-scale dataset is a bit lower than the ground-truth dataset.
In conclusion, the detection result shows that \system{} is effective for large-scale \target{} detection tasks. 
And the \target{} are extremely prevalent in the wild and Appstore. Even though the app stores have performed checks on their apps, we still find that there are some undetected \target{} (11\% of all apps) in AppStore.

\noindent \textbf{Performance.}
In addition, we evaluate the time consumption of our system. Overall, each app costs about \num{213} seconds on average, ranging from \num{9} seconds to $2,347$ seconds. We show the average time consumption of different app sizes in Figure~\ref{fig:time} (in Appendix B).
The \system{} time consumption is stable, where the fluctuation of time is only \num{67}s in the range \num{15}M-\num{57}M. 
The performance results show that \system{} is large-scale resilient.
And we further explain that our time consumption is much less than that of the dynamic detection method. Through our preliminary experiments, it takes more than 30 minutes for dynamic execution techniques to traverse an app. For large-scale \target{} detection, our approach is more lightweight.

\noindent \textbf{Measurement.}
We make a measurement of the \sg{}, including the number of transition pairs, the widgets, and the imprint tokens.

The investigation shows that the \sg{}s of \target{} and \legit{}s are significantly different in statistics. Specifically, the \target{} are the detected ones from the large-scale dataset, while the \legit{}s are the remaining ones. 
First, the average transition pairs in \legit{}s are twice as many as \target{} (\num{22} vs. \num{11}).
Second, the average number of widgets owned by \legit{}s is much higher than that of \target{} (\num{227} vs. \num{29}). 
Finally, turning to the network imprint. 
The average number of tokens owned by \legit{}s is also higher than that of \target{} (\num{211} vs. \num{121}).
The statistical result can  obviously proves our observation-I in Section~\ref{sec:motivation}, that \target{} are always single-purpose and different from \legit{}s.

\begin{tcolorbox}[size=title,colback=white]
\textbf{Answer RQ3:}
\system{} is large-size resilient and effective in large-scale detection.
The result suggests that \num{54}\% apps in the wild and \num{11}\% apps in the AppStore are \target{}.
\end{tcolorbox}
\section{Discussion}
\label{sec:discussion}

\noindent \textbf{Regional Differences and Commonalities.} 
The specific manifestations of \target{} are closely linked to local laws and regulations.
Although different regions/countries may have different regulations, our key observation is that there do exist differences between the non-compliant apps and the compliant ones (i.e., the normal apps) in a specific region~\cite{hongGamblnig2022, munyendo2022desperate,chen2021lifting}.
This suggests that our detection approach can be applied to other regions, though our datasets are collected from Asia. Based on our approach, different regions can establish their own datasets to accommodate their regulations.

\noindent \textbf{App Sketch Construction.}
As \sg{} is built using static analysis, apps may leverage app hardening techniques to disrupt static analysis.
In addition, even though the static analysis considers the major factors introduced by Android (\ie, multi-threading, lifecycle, and ICCs), apps may evade our analysis by invoking sensitive APIs via reflections and native libraries. 
On the other side, \sg{} uses network imprints to identify the network features. 
While the network imprints are highly malleable, they are still inaccurate compared to real remote resources.
In the future, we plan to incorporate dynamic analysis to deal with the app hardening techniques and capture real network remote resources. 

\noindent \textbf{Anti Adversary.}
Our model is trained by the ground-truth \target{} dataset.
To avoid our detection, adversaries may download our dataset and find patterns that are not covered.
For example, they can use specially designed app sketches to mislead our detection model. 
Since the dataset is collected manually by our team, the number of samples is limited which cannot prevent attacks against the dataset.
However, we claim that enterprises can supplement the dataset by themselves to improve the coverage, thereby improving the accuracy and avoiding attacks on the model coverage. 
In addition, \target{} classification categories can also be added according to the needs of enterprises.
\section{Related Work}
\noindent \textbf{Android UI Analysis.}
There are many studies focusing on Android UI analysis.
Azim et al.~\cite{azim2013targeted} presented the activity transition graph to model the Android UI transition.
Liu et al.~\cite{liu2022promal} combine machine learning and static analysis to complete the transition pairs.
Chen et al.~\cite{chen2019storydroid} took into account the inner class to complete the transition pairs.
Yang et al.~\cite{yang-icse15,yang-ase15} leveraged context-sensitive analysis of callback methods and developed a client analysis that builds a static GUI model and window transition graph.
PERUIM~\cite{li2016peruim} focused on the permissions used behind the widgets and connected the widgets with their handlers.
Gator~\cite{rountev2014static,yang2015static,yang2018static} modeled the WTG and the widget attributes by windows stack.
Frontmatter~\cite{kuznetsov2021all} implemented a lightweight static widget analysis tool, and linked the callback with buttons.
Our app sketch is built upon these static UI analysis techniques, and additionally adds new features (\ie, WebView, transition about the fragments, and navigation).

\noindent \textbf{Machine-Learning Based Detection}
Machine learning is widely used for anomaly detection. 
Previous studies~\cite{kim2018multimodal,ahmed2009using,li2018significant,yuan2014droid} leveraged the feature of API control flow and permissions to perform malware detection, since the malicious code is different from the benign code.
Besides, some studies leveraged anomaly pictures/texts and code to perform detection. 
AppIntent~\cite{yang2013appintent} used the taint analysis from the UIs to analyze the location that may cause sensitive data leakage.
AsDroid~\cite{huang2014asdroid} compares the GUI attribution and code behavior to find out the stealthy behavior against the normal GUI attribution.
IconIntent~\cite{xiao2019iconintent} leverages machine learning to automatically identify sensitive icon behaviors.
DeepIntent~\cite{xi2019deepintent} leverages machine learning to compare the difference between UI and code behavior to analyze the misusing icon.
SUPOR~\cite{huang2015supor} identified the privacy input based on the text, figures, and layout by machine learning.
In this study, we refer to the self-supervised algorithm and propose a novel algorithm to extract the \sg{} features and  identify the \target{}.

\section{Conclusion}
\label{sec:conclusion}
In recent years, the proliferation of UEware has become an emerging threat. This has seriously impacted the security of the mobile ecosystem and resulted in significant financial losses.
In this paper, we propose a novel \sg{} based approach to effectively and efficiently detect \target{}.
Specifically, it first statically builds precise \sg{} for \target{}, and then applies a graph-embedding based method to
represent UTG by properly correlating multi-dimensional attributes. After that, it leverages the self-supervised learning method to detect and classify \target{} based on the \sg{} feature similarity.
We have implemented a prototype system named \system{} to perform the evaluation.
The evaluation results show that \system{} is efficient and effective. It can detect \target{} more efficiently than the traditional approaches (with \num{98.22}\% detection F1-score and \num{98.97}\% classification accuracy). 
Furthermore, \system{} is large-size resilient and efficient for large-scale detection in the real world. By using \system{}, we found that \target{} are prevalent, \ie, \num{54}\% apps in the wild and \num{11}\% apps in the app stores are \target{}.
Our system could be used by authoritative agencies and app download platforms to effectively mitigate this threat while alleviating the human efforts required.

\section*{Data Availability Statement}
The prototype of the proposed \system{} system~\cite{DeUEDroid_System} is available from the corresponding author upon reasonable request.

\begin{acks}
We would like to thank all anonymous reviewers for their helpful suggestions
and comments to improve the paper. This work was supported by the National Key R\&D Program of China (No. 2022YFE0113200), the National Natural Science Foundation of China (No. U21A20464, 62172360, U21A20467, 62072395, and U20A20178), and the National Key Research and Development Program of China (No. 2020AAA0107705). The findings herein reflect the work and are solely the responsibility of the authors.
\end{acks}
\balance

\bibliographystyle{ACM-Reference-Format}
\bibliography{main}
\balance

\end{document}